\documentclass[times,11pt]{article}
\usepackage{times}
\usepackage{url}

\usepackage{epsfig}
\usepackage{algorithm}
\usepackage[noend]{algorithmic}
\usepackage{wrapfig}

\begin{document}


\title{Adaptive filter ordering in Spark}
\author{Nikodimos Nikolaidis\\
Aristotle University of Thessaloniki,\\ Greece \\
nikniknik@csd.auth.gr\\
\and
Anastasios Gounaris\\
Aristotle University of Thessaloniki,\\ Greece \\
gounaria@csd.auth.gr
}

\maketitle
\thispagestyle{empty}

\begin{abstract}
This report describes a technical methodology to render the Apache Spark execution engine adaptive. It presents the engineering solutions, which specifically target  to adaptively reorder predicates in data streams with evolving statistics.  The system extension developed is available as an open-source prototype. Indicative experimental results show its overhead and sensitivity to tuning parameters.
\end{abstract}

\section{Introduction}


Query processing technology heavily relies on two pillars, namely declarative query statement and cost-based optimization. The declarative manner in which database queries are submitted plays a key role in the wide-spread usage of database technology, whereas automated optimization allows non-expert users to develop particularly complex tasks without wondering about performance issues. Nowadays, both declarative (either SQL-like or operator-based) query statement and (either heuristic or cost-based) optimization have been leveraged in the Apache Spark execution engine.  

However, when queries are long running or even continuous, a static execution plan is typically sub-optimal due to the evolving data characteristics. To address this problem, several adaptive query processing (AQP) techniques have been developed in the 2000s \cite{BNCOD2002,Deshpande07} that have been incorporated in complete system prototypes, e.g., \cite{TelegraphCQ} and modern DBMSs. One of the first problems that AQP tackled was adaptive ordering of filter operators to reflect the volatile data-dependent cost and selectivity values of such operators, e.g. \cite{357BMM+04,Avnur00}.

Although AQP solutions for distributed settings exist \cite{GounarisTM13}, these solutions have not been transferred to a massively parallel setting, such as that of typical Spark or Flink executions. Current adaptive solutions for such engines are restricted to issues, such as runtime choice of the physical operator implementation \cite{KBC+18}. Our work shows how to incorporate adaptive filter ordering in Spark.
To this end, several requirements need to be met so that the solution is:
\begin{enumerate}
\item \emph{Efficient:} the extension should be lightweight and not incur high overhead; we discuss this in Sec. \ref{technical} and \ref{exps}.

\item \emph{Effective:} for long-running queries on volatile data, it should yield tangible performance improvements; we provide concrete examples in Sec. \ref{exps}.

\item \emph{Publicly available and pluggable:} our solution is provided online from \\ \url{https://spark-packages.org/package/kikniknik/spark-adaptive_filtering} and \url{https://github.com/kikniknik/spark-adaptive_filtering}. It can be employed by third parties by just including a provided library.

\item \emph{Configurable:} in the next section, we explain the main parameters and we evaluate them in Sec. \ref{exps}.

\item \emph{Extensible:} we briefly discuss extensibility in Sec. \ref{discussion}.

\end{enumerate}

\section{Technical Details}
\label{technical}

Suppose we want to analyze structured server logs in a data-frame \texttt{df} and focus on events occurring between 07:00 and 16:00 when all memory, CPU and network resources are relatively stressed. Then, an example of expressing such a task in Spark would seem like the following:
\texttt{df.filter(hour >  7 \&\& hour < 16 \&\& memoryUsage > 60 \&\& cpuUsage > 60 \&\& networkUsage  > 30)}. The default implementation of the \texttt{filter} (i.e., select)
physical operator  in Spark does not include a mechanism to specify an order in which predicates will be evaluated  in selection queries neither in advance nor during query execution. The execution order is as specified in the statement. In the above example, the first hour filter will be evaluated first, followed by the second filter on hour, and so on.
 As such, the  decision of selection ordering rests with the user.

As already explained, in datasets with evolving data characteristics, continuously revising the execution order is
better than having a static execution plan.  Therefore,  our goal is to constantly specify a good adaptive order based on how selective and expensive each predicate is. In order to achieve that, the \texttt{filter} operator has to be extended to support (i) predicate evaluation monitoring and statistics collection; and (ii) on-the-fly predicate re-ordering. In this context, many strategies may be applied, differing in which metrics are collected and how predicates are re-ordered. Our choices are described below.

\subsection{Description of the approach}

According to the standard database theory, filters should be ordered using two metrics per filter, namely
\textit{selectivity (s)} and \textit{cost (c)}. The $rank = \frac{c}{1-s}$ metric combines them and it can be easily proven that ordering filters by their rank values in ascending order minimizes the sum of the individual costs, and thus the resource consumption.  

The first step towards adaptive filter ordering is to monitor the potentially evolving cost and selectivity values for each filter inducing as low overhead as possible.
We sample one row in every {\tt collectRate} rows, on
 which every predicate is evaluated and monitored - no matter if a precedent predicate is satisfied or not. We want to monitor a relatively small subset of all rows to avoid a high overhead cost but we avoid sampling by pseudo-random generators because they carry an additional cost. By evaluating all predicates in monitored rows, we avoid bias from correlated values in line with the approach in \cite{357BMM+04}.

The metrics that are collected for each monitored row consist of two arrays of size equal to the number of predicates, {\tt numCut} and {\tt cost}. Each position of these arrays corresponds to a predicate in the initial order given by the user. {\tt numCut} holds the number of rows that did not satisfy each predicate and {\tt cost} holds the total execution time dedicated to each predicate for its evaluation, which is calculated through the system's clock. Having these metrics and the total number of monitored rows, we then calculate the \textit{selectivities} and \textit{costs} of the predicates.

We split the execution of a stream or a large dataset in epochs (or phases), specified in dataframe rows.
Each epoch corresponds to the processing of {\tt calculateRate} rows. At the beginning of each phase, predicate ranks are (re-)calculated and predicates are sorted in ascending order based on their ranks.  The specified order is kept for the whole epoch duration.

For the calculation of ranks, the average costs are normalized to a range $[0,1]$, so that they are in the same scale with selectivities. Thus, the rank of predicate $i$ at moment $t$ is defined as $rank_i^{(t)} = \frac{nc_i^{(t)}}{1-s_i^{(t)}}$, where $nc$ is the normalized average cost. More importantly, we add a factor of {\tt momentum} ($m$) that preserves a percentage of previous rank. Overall, we consider an adjusted rank with the help of a first-order difference equation:
$$ adj\_rank_i^{(t)} = (1-m) \cdot rank_i^{(t)} + m \cdot adj\_rank_i^{(t-1)} $$
The use of a momentum helps in the stability of the approach  and in avoiding temporary fluctuations.

Table \ref{table:doConsumAdaptiveConf} summarizes the main configuration parameters along with the default values in our tests.  As shown, the default sampling rate is 0.1\% and the ordering is revised after processing 1 million tuples.
\begin{table}[tb!]
	\begin{center}
\scriptsize
		\begin{tabular}{ | l | l | l | }
			\hline
			\textbf{Name} & \textbf{Default} & \textbf{Description} \\
			\hline
			{\tt collectRate} & $1000$ & Statistics collect rate (in rows) \\ \hline
			{\tt calculateRate} & $1000000$ & Ranks calculation rate (in rows) \\ \hline
			{\tt momentum} & $0.3$ & Past preservation factor \\ \hline
		\end{tabular}

\caption{Configuration options for the adaptive filter operator in Spark}
\label{table:doConsumAdaptiveConf}
	\end{center}
\end{table}

\subsection{Implementation in Spark}

As of 2.0 version of Apache Spark, some operators including \texttt{filter}, support code generation \cite{armbrust2015spark}. Following this principle, we designed the adaptive \texttt{filter} operator, so that it generates the necessary code for filtering inside an iterator's {\tt processNext} method. The output of that method is used by the parental node of execution plan tree.

In contrast to Spark's implementation, where predicates are evaluated statically (in the order given by the user) inside the {\tt processNext} method, we keep an array of a specific permutation of predicates (given by best order based on ranks) and we evaluate the predicates according to that permutation. To achieve this, we employ, a  function with a switch statement and one case for every predicate. Then, inside the {\tt processNext} method, this function is called for every predicate in the order given by the permutation array.



The \texttt{filter} operator runs in a distributed manner inside \emph{Executors}, and each partition of data is processed by a different thread of the executor JVM, constituting a \emph{task}. A decision needs to be drawn regarding the scope and lifetime of (adjusted) ranks, which are using information from the past through momentum, as we mentioned earlier. Metadata may live independently in tasks, or in executors or even be centralized in the driver. Tasks have a relatively small lifetime because of quick partition exhaustion, so if metadata is kept on a \emph{per-task} basis, this would mean that
 ranks will have a small lifetime and they will need to be initialized in each task, which does not let enough information to be aggregated, so that a good picture of the data is built.
In a \textit{centralized} policy, some extra traffic will be incurred to the network, causing waiting times and communication issues. The \textit{per-executor} policy overcomes the disadvantages of the two afore-mentioned policies. Ranks live as long as the job is running and no data is required to be transferred through the network.
Additionally, in case of heterogeneous data, autonomous executors can have their own predicate ranks and the imposed orders can follow the local data properties, which renders the technique more adaptive.

We implemented the \textit{per-executor} policy by using static variables for permutation and ranks, which means that they are global in the executor JVM. Nevertheless, tasks autonomously collect the metrics ({\tt numCut}, {\tt cost}) and they update global ranks based on what they collected in their last phase.
It might be the case that multiple tasks, while processing separate partitions running inside the same executor, may attempt to alter the execution order concurrently. In that case, through the usage of a simple lock, only one task is permitted to alter the order in a single epoch. Non-permitted updates are deferred to the next epoch keeping the collected metrics.


\emph{How the code can be used.} Third parties can employ our solution through Catalyst, where the new code is seen as an extension.
The implementation consists of one package. Apart from the adaptive \texttt{filter} physical operator, a Catalyst {\tt SparkStrategy} class is provided where logical filters are transformed to physical adaptive filters instead of default Filters. This class can be added as an extension to Spark through {\tt SparkSessionExtensions}, so that whenever a filter is included in a query, SparkPlan will employ Adaptive Filters instead of normal ones.

\section{Experiments}
\label{exps}

In the following experiments, we use a
synthetic dataset of 75M rows and 3 attributes of different types, namely date, integer, and string; all attribute values follow a normal  distribution.

\subsection{All possible permutations: Adaptive vs Non Adaptive}


\begin{figure}
\centering
\includegraphics[width=\columnwidth]{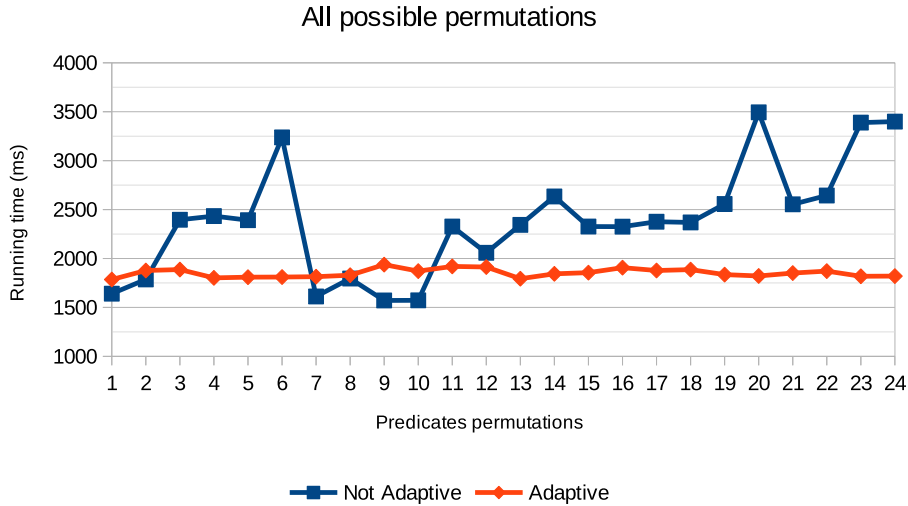}
\caption{Adaptive vs Non-adaptive for all possible static orderings}
\label{fig:permutations}
\end{figure}

In this experiment, we focus on effectiveness (performance improvements) and efficiency (low overhead). We employ four filter conditions (2 on the integer attribute and one on the date and string ones). The overall selectivity of the query is 4.51\%. It is trivial to construct cases where the different filter orderings yield performance differences of several orders of magnitude; here, we examine a more realistic case, where the best and the worst performing orderings differ by 2.3X.

Overall, there are 24 possible static orderings; we evaluate each one of them against the adaptive approach.
The results are shown in Figure \ref{fig:permutations}. There are two main observations: firstly, our approach is always very close to the optimal statically defined orderings yielding improvements of more than 2X in this setting; secondly, the overhead of the approach is low and the solution is robust to the initial ordering defined by the user.

\subsection{Sensitivity analysis}

\begin{figure}[tb!]
\centering
\includegraphics[width=\columnwidth]{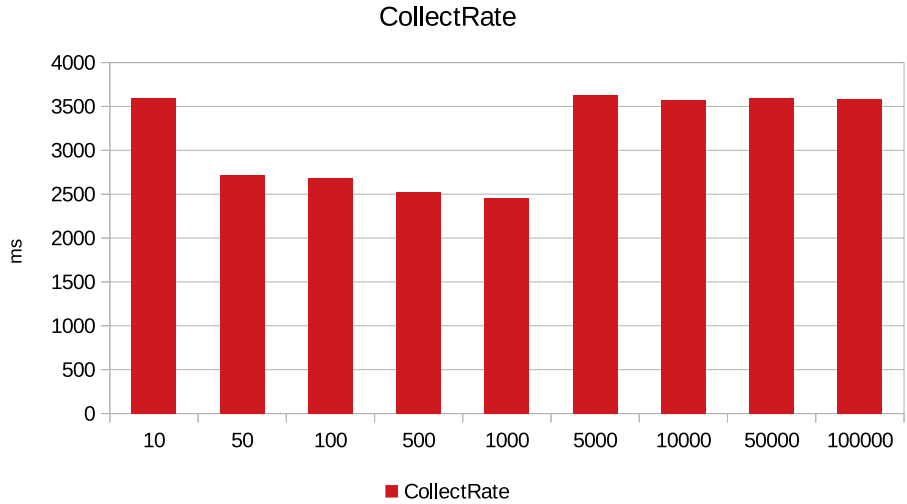}
\caption{Impact of {\tt collectRate}}
\label{fig:collectRate}
\end{figure}
\begin{figure}[tb!]
\centering
\includegraphics[width=\columnwidth]{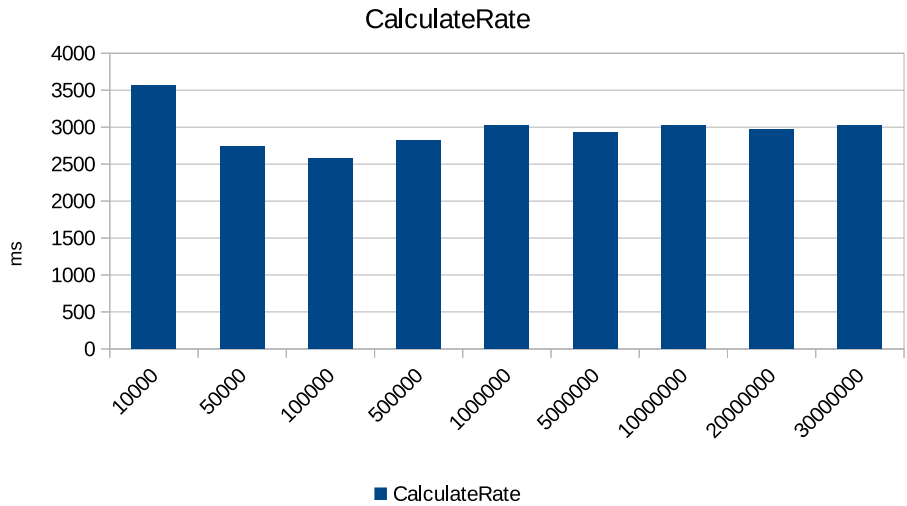}
\caption{Impact of {\tt calculateRate}}
\label{fig:calculateRate}
\end{figure}
\begin{figure}[tb!]
\centering
\includegraphics[width=\columnwidth]{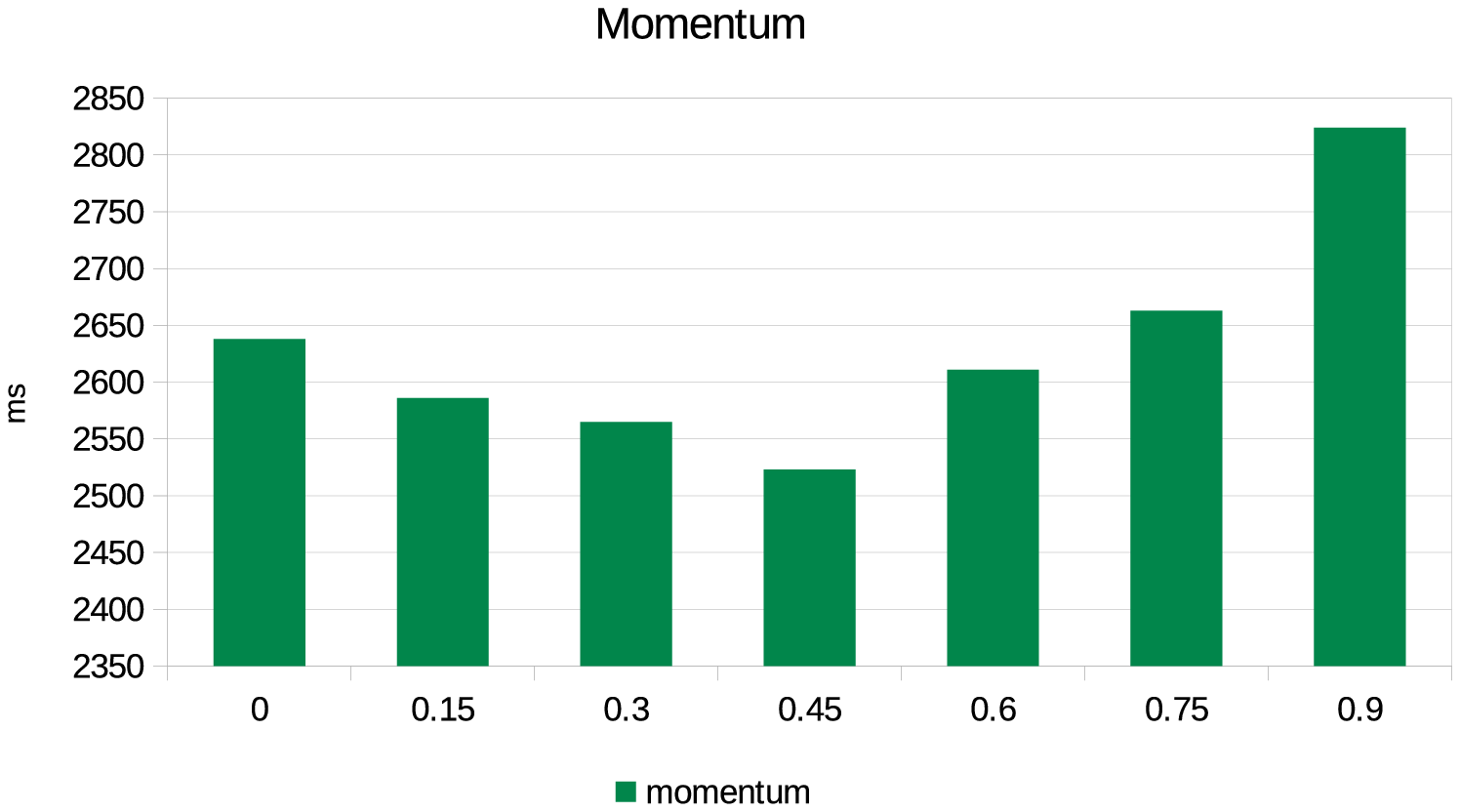}
\caption{Impact of {\tt momentum}}
\label{fig:momentum}
\end{figure}

In this experiment, we focus on the impact of the configuration parameters of {\tt collectRate}, {\tt calculateRate} and {\tt momentum}.
For the differences to be noticeable, we slightly modify the conditions, so that 16.14\% of the tuples pass all four filters. The results are shown in Figures \ref{fig:collectRate}, \ref{fig:calculateRate} and \ref{fig:momentum}. As expected, both very low and very high values lead to performance degradation; by contrast, middle values achieve the best trade-off between low overhead by not being too aggressive and not being very slow to adapt.

\section{Discussion}
\label{discussion}

This report aims to describe an adaptive query processing technique for Spark that is the first one, to the best of our knowledge, that modifies the execution plan on the fly. In its current form, it supports only filters but can be extended in several ways. For example, a policy similar to A-greedy \cite{357BMM+04} or even eddies \cite{Avnur00} can be implemented.
Supporting joins, in line with the techniques in \cite{Deshpande07} is also a promising direction. In parallel with modifying the execution plan, adaptive techniques for the optimal configuration of the parameters can be adopted, e.g., \cite{GounarisYSD08b,GounarisYSD08}.

\end{document}